# A large moment antiferromagnetic order in overdoped high-$T_c$ superconductor $^{154}$SmFeAsO$_{1-x}$D$_x$

Soshi Iimura[1], Hiroshi Okanishi[1], Satoru Matsuishi[2], Haruhiro Hiraka[3], Takashi Honda[4], Kazutaka Ikeda[4], Thomas C. Hansen[5], Toshiya Otomo[4,6], and Hideo Hosono[1, 2, *]

[1]Laboratory for Materials and Structures, Tokyo Institute of Technology, Yokohama 226-8503, Japan

[2]Research Center for Element Strategy, Tokyo Institute of Technology, Yokohama 226-8503, Japan

[3]Graduate School of Science and Engineering, Ibaraki University, Mito, Ibaraki 310-8512, Japan

[4]Institute of Materials Structure Science, High Energy Accelerator Research Organization (KEK), Tsukuba, Ibaraki 305-0801, Japan

[5]Institut Laue-Langevin, 6 rue Jules Horowitz, BP 156, 38042 Grenoble Cedex 9, France

[6]Department of Materials Structure Science, The Graduate University for Advanced Studies, Tsukuba, Ibaraki 305-0801, Japan

**Abstract**

In iron-based superconductors, high critical temperature ($T_c$) superconductivity over 50 K has only been accomplished in electron-doped *hRE*FeAsO (*hRE* = heavy rare earth (*RE*) element). While *hRE*FeAsO has the highest bulk $T_c$ (58 K), progress in understanding its physical properties has been relatively slow due to difficulties in achieving high concentration electron-doping and carrying out neutron-experiments. Here, we present a systematic neutron powder diffraction (NPD) study of $^{154}$SmFeAsO$_{1-x}$D$_x$, and the discovery of a new long-range antiferromagnetic ordering with $x \geq 0.56$ (AFM2) accompanying a structural transition from tetragonal to orthorhombic. Surprisingly, the Fe magnetic moment in AFM2 reaches a magnitude of 2.73 $\mu_B$/Fe, which is the largest in all non-doped iron pnictides and chalcogenides. Theoretical calculations suggest that the AFM2 phase originates in kinetic frustration of the Fe-3$d_{xy}$ orbital, in which the nearest neighbor hopping parameter becomes zero. The unique phase diagram, *i. e.*, highest-$T_c$ superconducting phase is adjacent to the strongly correlated phase in electron-overdoped regime, yields important clues to the unconventional origins of superconductivity.

**Significance statement**

Using NPD measurements on the bulk highest $T_c$ superconductor $^{154}$SmFeAsO$_{1-x}$D$_x$, we discovered a new AFM phase in the electron-overdoped regime, $x \geq 0.56$ (AFM2). The magnetic moment on Fe in AFM2 reaches 2.73 $\mu_B$/Fe, which is the largest in all the non-doped iron-based antiferromagnets reported so far. Our theoretical calculations reveal that the AFM2 phase in SmFeAsO$_{1-x}$H$_x$ originates in the kinetic frustration of the Fe-3$d_{xy}$ orbital, in which the Fe-3$d_{xy}$ nearest neighbor hopping parameter becomes zero. The unique phase diagram, *i.e.*, the highest-$T_c$ superconducting phase is adjacent to the strongly electron-correlated phase in heavily electron-doped regime (not nondoped regime), yields important clues to the unconventional origins of superconductivity.

¥body

**Introduction**

Carrier doping is a critical parameter that governs the electronic correlations and ground states in high-$T_c$ superconductors. Electronic phase diagrams depicting the evolution of the ground state with doping level not only deepen our understanding of the mechanism of superconductivity, but help us extract common trends across different superconducting materials. Until recently, two electronic phases were believed to play a vital role in the iron-based superconductivity, *i. e.*, a stripe or double-stripe-type antiferromagnetic (AFM) ordering at a non-doped Fe-$3d^6$ state and a Mott insulator at a hole-doped Fe-$3d^5$ state[1–3]. The former phase is observed at ambient pressure for the arsenides and telluride, while under high pressures for the selenides, and the AFM fluctuations are a promising candidate for the pairing glue leading to the superconductivity[4,5]. On the other hand, the latter picture, confirmed recently in effectively hole-doped Na(Fe$^{3+}_{0.5}$Cu$^+_{0.5}$)As[6], explains the asymmetric response of the $T_c$ and electron correlation to hole- and electron-doping, where the hole-doping (electron-doping) into the Fe-$3d^6$ state tends to enhance (reduce) the $T_c$ and the degree of electron correlation, since the system approaches (goes away from) the half-filled Fe-$3d^5$ state[2].

However, these two prevailing scenarios were challenged recently by observations of the behavior of heavily electron-doped FeSe (11-type) and *RE*FeAsO (1111-type), where the $T_c$ and electron correlation strength are enhanced rather than weakened with electron-doping. In the first of these systems, a quite high $T_c$ in the range of 35-41 K was observed for the electron-doped FeSe using an electric double-layer transistor, which climbs to 46-48 K for the potassium-coated bulk FeSe, and rises above 100 K for FeSe-monolayer[6–10]. Remarkably, in the overdoped regime of the potassium-coated bulk FeSe, an electronic-correlation-driven insulating phase is developed, where an AFM ordering is also proposed[8]. The *RE*FeAsO phases are AFM metals at low temperature. The electron-doping via hydride-ion-substitution into the oxygen-site results in the suppression of AFM (AFM1) ordering and subsequently the emergence of a superconducting phase for the electron-doping range of $0.05 \leq e^-/\text{Fe} \leq 0.45$[11,12]. The optimum $T_c$, which depends strikingly on the *RE* size, is 36 K for *RE* = La, and 56 K for Sm or Gd, consistent with the observation that the application of physical pressure to La-1111 can enhance its $T_c$ from 36 K to 52 K[12,13]. In the overdoped regime of La-1111 (electron-doping levels $\geq 0.40e^-$/Fe), another long-range AFM ordering (AFM2) was observed by the NPD and muon spin relaxation techniques[14]. The relatively large Fe magnetic moment in the AFM2 phase (1.21$\mu_B$/Fe) and recent theoretical calculations using density-functional theory plus dynamical mean-field theory (DFT+DMFT) suggest that pronounced correlation occurs in the overdoped regime[15].

In spite of the fascinating properties of electron-doped *hRE*FeAsO, its phase diagram, especially its magnetic aspects, is not well understood probably due to the prohibitively large neutron absorption cross sections of Sm and Gd, which are two or three orders of magnitude larger than that of La. Even for the non-doped case, the structure of the magnetic ground state and the magnitudes of magnetic moments remain unknown, though muon spin rotation and magnetic x-ray scattering confirm the existence of AFM order in SmFeAsO[16,17]. Here, we have synthesized electron-overdoped SmFeAsO$_{1-x}$H$_x$ with $x$ = 0.51, 0.59, 0.67, 0.76, and 0.82 under 1573 K and 5GPa. To reduce the large neutron absorbance of natural Sm, the isotopically substituted $^{154}$SmFeAsO$_{1-x}$D$_x$ samples were also prepared at $x$ = 0, 0.56, 0.67, and 0.73 to allow NPD measurements. All samples showed metallic conductivity and no superconducting transition above $T$ = 2 K (the detailed synthesis procedures, the $x$ dependence of the structural parameters, and the results of resistivity measurements are provided in the SI Appendix, Fig. S1, S6, and S7).

S. Iimura *et al.*,

## Results and Discussion

In **Fig. 1A** we show the calculated and observed NPD patterns of $^{154}$SmFeAsO$_{0.27}$D$_{0.73}$ collected on a high resolution diffractometer NOVA at $T = 10$ K. An excellent fit to the data was achieved using space group *Aem*2, the same as that adopted by LaFeAsO$_{0.49}$H$_{0.51}$ at low temperatures[14]. The structural transition from the tetragonal *P*4/*nmm* to the orthorhombic *Aem*2 space group is obvious from the NPD patterns: below $T = 90$ K (light blue line), the tetragonal 220 reflection splits asymmetrically into the orthorhombic 040 and 004 reflections, whereas the tetragonal 200 and 001 reflections are unchanged on cooling (**Fig. 1B**), indicating a rhombic distortion in *ab*-plane of the *P*4/*nmm* phase. The evolution of the refined lattice constants as a function of temperature is shown in **Fig. 1C** and **1D**. The *ab* splitting and an upturn of the *c*-axis length take place at around $T = 90$ K, consistent with the detection of second-order phase transitions in our heat capacity ($C_p$) measurements at $T = 88, 89, 90, 86,$ and 89 K for SmFeAsO$_{1-x}$H$_x$ samples with $x = 0.51$, 0.59, 0.67, 0.76, and 0.82, respectively, and $T = 86, 86,$ and 86 K for $^{154}$SmFeAsO$_{1-x}$D$_x$ samples with $x = 0.56, 0.67,$ and 0.73, respectively. (see the SI Appendix, Fig.S5). Although the structural transition temperature ($T_s$) is likely to be doping-independent, the orthorhombicity, $\delta = (b_O - c_O)/(b_O + c_O)$, at $T = 10$ K monotonically increases with $x$: $\delta = 0.24, 0.37,$ and 0.40 for the samples with $x = 0.56, 0.67,$ and 0.73, respectively.

In order to discuss the static magnetic ordering occurring in $^{154}$SmFeAsO and $^{154}$SmFeAsO$_{1-x}$D$_x$, we turn to the results of high-intensity NPD experiments carried out on D20. NPD profiles measured on $^{154}$SmFeAsO at $T = 2, 6,$ and 170 K are shown in **Fig.2A**. Compared with the profile at $T = 170$ K, additional peaks appear in the pattern at $T = 6$ K at $2\theta \sim 27, 36$ and 59 deg., indicating that a paramagnetic-antiferromagnetic transition occurs in the range $6 < T < 170$ K. On further cooling to $T = 2$ K, changes in the intensity of peaks at $2\theta \sim 35.5$ and 40 deg. indicate the occurrence of another AFM ordering. From the temperature dependences of areas of these peaks (**Fig.2B** and **2C**), the two AFM transition temperatures are determined to be 137(2) K and 5.4(9) K, which are interpreted as the Neel temperatures of the spins of the Fe- ($T_N^{Fe}$) and Sm-sublattices ($T_N^{Sm}$), respectively. The magnetic structure determined for $^{154}$SmFeAsO is illustrated in **Fig. 2D** and **2E**. The AFM structure of Fe is described by a propagation vector *k* of (0, 1, 0.5); namely, the moments of the Fe atoms are aligned antiferromagnetically along the *b* and *c* axes, and ferromagnetically along the *a* axis. Meanwhile, the propagation vector for the AFM structure of the Sm atoms is *k* = (0, 0, 0) where the Sm moments are aligned antiferromagnetically along the *c* axes, and ferromagnetically within the *ab* plane. The refined magnitudes of the Fe and Sm moments are 0.66(5)μ$_B$/Fe and 0.62(3)μ$_B$/Sm, respectively. Hereafter, we will refer to the AFM phase of this *x* = 0 material as AFM1.

A thermodiffractogram of $^{154}$SmFeAsO$_{0.27}$D$_{0.73}$ over the temperature range 2 ≤ *T* ≤ 70 K clearly tracks the growth of AFM ordering upon cooling below *T* ~ 60 K, especially in 20 ≤ 2θ ≤ 30 deg. window (**Fig.2F**). Fitting the temperature variation in the peak areas at 2θ ~ 20 deg. to a power-law function, $A(T_{N1} - T)^{2\beta} + B$, we obtain $T_{N1}$ = 51(4) K and a *β* value of 0.33(3) which is close to that expected for a three-dimensional Ising model (*β* ≈ 0.325) (**Fig.2G**)[18]. In addition, the intensity of the peak at 2θ ~ 26 deg. shows an abrupt jump at *T* = 30-35 K ($T_{N2}$) on heating, above which a new peak appears at 2θ ~ 36.5 deg. (**Fig.2H**). The two transitions temperature designated as $T_{N1}$ and $T_{N2}$ coincide well with the two broad peaks observed in our $C_p$ measurements at 57 and 33 K, indicating that the anomalies at $T_{N1}$ and $T_{N2}$ correspond to some form of magnetic phase transitions. From a comprehensive search for the *k*-vector at *T* = 2 and 39 K within the Brillouin zone of the *A*-centered orthorhombic cell, the magnetic Bragg peaks at *T* = 2 K were perfectly indexed as an incommensurate single-*k* structure with *k* = (0, 0.773, 0) for both the Fe and Sm-sites (**Fig.2I**), while those at *T* = 39 K were as fit to an incommensurate *k*-vector of (0, 0.800, 0) for the Fe and Sm-sites and a commensurate *k*-vector of (0.5, 1, 0) for the Fe-site. In other words, the magnetic structure at *T* = 39 K is a multi-*k* structure (**Fig.2J**)[19]. The single-*k* incommensurate structure of $^{154}$SmFeAsO$_{0.27}$D$_{0.73}$ (IC-AFM2) at *T* = 2 K is displayed in **Fig.2K** and **Fig.2L**. The spin orientations on the Fe-sites almost seems to follow a longitudinal spin-density wave, where the spins are parallel to the incommensurate *k*-vector. Unlike the independent AFM orderings of Fe and Sm's spins at $T_N^{Fe}$ = 137(2) K and $T_N^{Sm}$ = 5.4(9) K in the AFM1, respectively, both the spins simultaneously order at $T = T_{N1}$ in the AFM2, indicating a strong coupling between them (The details regarding the magnetic structure determination and the structure of the multi-*k* AFM (C+IC-AFM) phase are provided in the SI Appendix Fig.S2-4). The most remarkable feature of the IC-AFM2 phase is that the magnitude of the Fe magnetic moment reaches as high as 2.73(6) $\mu_B$/Fe. Although iron pnictides have been considered to be less correlated compounds than chalcogenides due to their strong covalency, this magnetic moment is larger than those of all the non-doped iron-based compounds, including pnictides and chalcogenides, and is rather comparable to those of electron-doped, Fe-vacancy-ordered $A_z\text{Fe}_{2-\delta}\text{Se}_2$ (3.2-3.4 $\mu_B$/Fe, $A$ = K, Rb, Cs, and Tl) materials[1,20].

**Figure 3A** compares the low temperature NPD profiles of $^{154}\text{SmFeAsO}_{1-x}\text{D}_x$ with $x$ = 0.56, 0.67, and 0.73 collected at $T = 10$ K. As $x$ is increased, the intensities of the magnetic Bragg peaks are enhanced, and their positions shift to higher $d$ values. The refined magnetic moments and the *k*-vectors for this series are summarized in **Fig. 3B** and **Fig. 3C**, respectively. The magnetic moments on the Fe- and Sm-sites grow sharply with increasing $x$, from 0.77(13) $\mu_B$/Fe and 0.40(16) $\mu_B$/Sm at $x$ = 0.56 to 2.77(10) $\mu_B$/Fe and 1.20(11) $\mu_B$/Sm at $x$ = 0.73, while the $y$-component of the *k*-vector ($k_y$) decreases linearly with $x$ from 0.869(6) at $x$ = 0.56 to 0.7730(4) at $x$ = 0.73. The $x$ dependence of the magnetic structures is illustrated in **Fig. 3D-3F**. The AFM structure of Fe sublattice at $x$ = 0.56 is a transverse spin-density wave, where the spins are perpendicular to the incommensurate *k*-vector. With increasing $x$, the angle denoted by $\Theta$ in **Fig. 3D-3F** gradually shrinks from 90° for $x$ = 0.56 to 11° for $x$ = 0.73, thus resulting in the longitudinal-spin-density wave at $x$ = 0.73. The Sm's spins are also likely to change their direction of alignment from the *c*-axis to the *b*-axis with doping.

The obtained transition temperatures are summarized in the form of a phase diagram in **Fig. 4A**. As with most iron pnictides, the ground state of the non-doped sample is a orthorhombic stripe-type AFM ordering with a small magnetic moment less than 1$\mu_B$/Fe (0.66(5) $\mu_B$/Fe). A dome-like superconducting phase appears after the AFM1 is suppressed at $x \sim 0.05$. In the overdoped regime, another orthorhombic AFM phase develops, in which the $T_{N1}$, $T_{N2}$, and the magnetic moments on the Fe- and Sm-sites increase with *x*. Considering that the isostructural LaFeAsO$_{1-x}$H$_x$ also exhibits the orthorhombic AFM2 phase in the overdoped regime, the bipartite AFM phases sandwiching the superconducting phase appear to be a common characteristic of the electron-doped 1111-type system[14]. Unlike the commensurate AFM2 phase in electron-overdoped LaFeAsO$_{0.49}$H$_{0.51}$, the AFM2 in $^{154}$SmFeAsO$_{1-x}$D$_x$ is incommensurate. This is perhaps due to the strong coupling between the spins of the Fe and Sm sublattices. Indeed, the IC-AFM phase emerges with simultaneous ordering on the Fe- and Sm-sites, whereas the C-AFM phase appears only on the Fe-sites.

Here, we compare the physical properties in the electron-overdoped regime of SmFeAsO with those in other iron-based superconductors. So far, heavy electron-doping has been considered to reduce the electron correlation strength, leading to a Fermi liquid in the overdoped regime, since the system goes away from a half-filled Fe-3$d^5$ state[2]. For example, the effective mass in the BaFe$_2$As$_2$ is strongly differentiated among Fe-3$d$ orbitals, where, in particular, the mass of a $d_{xy}$ electron is markedly reduced by electron-doping via Co-substitution onto the Fe-site[21]. Angle-resolved photoemission spectroscopy (ARPES) investigations of LiFe$_{1-x}$Co$_x$As and NaFe$_{1-x}$Co$_x$As also support this scenario in that the width of the $d_{xy}$-based band is significantly broadened with increasing Co content[20]. In the same manner, the overdoped *RE*Fe$_{1-x}$Co$_x$AsO and Fe$_{1-x}$Co$_x$Se show Fermi liquid-like behavior as is confirmed by transport measurements[22–24].

In sharp contrast with these direct-doping cases where electrons are supplied by an aliovalent cation substituted on the Fe-site, heavy electron-doping via indirect means, such as the hydride substitution in *RE*FeAsO$_{1-x}$H$_x$ and K-coating for FeSe (K$_y$FeSe), effectively enhances the electron correlation strength[9,15]. Indeed, even for the canonical BaFe$_2$As$_2$ system, indirect electron-doping via La$^{3+}$-substitution onto the Ba$^{2+}$-site doesn't lead to a quadratic resistivity $\rho \sim T^2$ in the overdoped regime[25]. These data may suggest that the evolution of electron correlation and the resultant physical properties are governed by not only the amount of electron, but also by the doping-site, *e.g.*, direct or indirect, or at least the concomitant structural changes that result from these different types of doping.

To clarify the effect of indirect electron-doping on the electronic structures of SmFeAsO and FeSe, we show the phase diagram of $K_yFeSe$ and the doping dependence of the calculated nearest neighbor hopping parameters of the Fe-3$d_{xy}$ orbitals ($t_1^{dxy}$) in the two systems in **Figure 4B-D** (the calculation method is described in detail in the SI Appendix.). **Fig.4C** and **4D** depict the four sets of $t_1^{dxy}$ values calculated using models that consider *i*) both the effects of Indirect-Doping and Structural changes (IDS), (*ii*) only Indirect-Doping (o-ID), (*iii*) only Structural changes (o-S), and (*iv*) only Direct-Doping via Co-substitution (o-DD). For the IDS model of SmFeAsO$_{1-x}$H$_x$, we performed two calculations: one carried out on a supercell (IDS_SC) containing hydrogen as an electron dopant, while the other employed the virtual crystal approximation (IDS_VCA) treating the oxygen as a virtual atom with a fractional nuclear charge ($Z = 8 + x$). In both compounds, the $t_1^{dxy}$ parameters for the IDS models are strongly reduced with increased doping and then approach zero in the AFM2 regime of $x > 0.5$ for SmFeAsO$_{1-x}$H$_x$ and the insulating regime of $y \sim 0.2$ for K$_y$FeSe. A similar rapid decrease is obtained in the o-ID model, whereas the $t_1^{dxy}$ values in the o-S and o-DD models show smaller sensitivity to changes in the structure and the Co-content, respectively, strongly indicating that the indirect electron-doping is the primary factor affecting $t_1^{dxy}$.

Previously, such a strong reduction in the $t_1^{dxy}$ parameters of iron-based superconductors was attributed to the longer Fe-anion and the shorter Fe-Fe bond lengths, which respectively reduce the indirect contributions to $t_1^{dxy}$ and increase the direct one, which cancel each other due to their opposite signs[20,26,27]. This structural changes and the resultant $t_1^{dxy}$ parameter successfully explain the trends in many physical properties of these system, *e. g.*, the largest Fe moment of 1.9 $\mu_B$/Fe being observed for FeTe with its long Fe-Te bond and suppressed $t_1^{dxy}$ value of 0.05 eV[20]. Indeed, the Fe-As and Fe-Fe bond lengths in SmFeAsO$_{1-x}$H$_x$ are elongated and is shortened respectively, as *x* is increased (see the SI Appendix Fig.S7). However, what we have revealed here is that the indirect electron-doping is much more significant than those of such structural effects. Therefore, we conclude that electron localization in the overdoped regimes of SmFeAsO$_{1-x}$H$_x$ and K$_y$FeSe arise from kinetic frustration induced by indirect electron doping.

We have discovered the orthorhombic AFM2 phase in the electron-overdoped regime of $^{154}$SmFeAsO$_{1-x}$D$_x$ with $x \geq 0.56$. The magnetic structures of both the Fe and Sm in the AFM2 (at $x \geq 0.56$) and AFM1 ($0 \leq x \leq 0.05$) phases were determined to be unique incommensurate and the conventional stripe type structures, respectively. The large Fe magnetic moment of 2.73 $\mu_B$/Fe in AFM2 phase is not only 4 times larger than that in AFM1 (0.66$\mu_B$/Fe) but is also the largest in all the non-doped iron-based superconductors. Our theoretical calculations reveal that the nearest neighbor hopping parameter of the Fe-$3d_{xy}$ orbital is strongly reduced by the indirect electron doping and becomes zero in the overdoped regime. This kinetic frustration induced by the indirect electron-doping strikes a strong contrast with the less-correlated Fermi liquid state in the electron-overdoped regime of Co-substituted iron-based superconductors. The unique phase diagram of $^{154}$SmFeAsO$_{1-x}$D$_x$ sheds light on the impact of the quantum fluctuations derived from the strongly correlated phase in electron-overdoped regime on the highest-$T_c$ superconductivity.

## Acknowledgement

We thank Professor Daniel C. Fredrickson for checking our manuscript for grammar and spelling mistakes. This study was supported by the MEXT Element Strategy Initiative Project to form a research core and Grant-in-Aid for Young Scientists (B) (Grant No. 26800182) from JSPS. The neutron-experiments were performed at BL21-NOVA on J-PARC (Proposal No. 2014S06) and D20 on Institut Laue-Langevin (Proposal No. INDU-106).

## Author contributions

S.I. and H.Hosono conceived the study. S.I., H.Hiraka, T.H., K.I., T.C.H. and T. O. carried out the neutron-experiments. S.I., H.O., and S.M. synthesized the samples. S.I. and H.Hosono wrote the manuscript. All the authors discussed the results and the manuscript.

S. Iimura *et al.*,

**Figures**

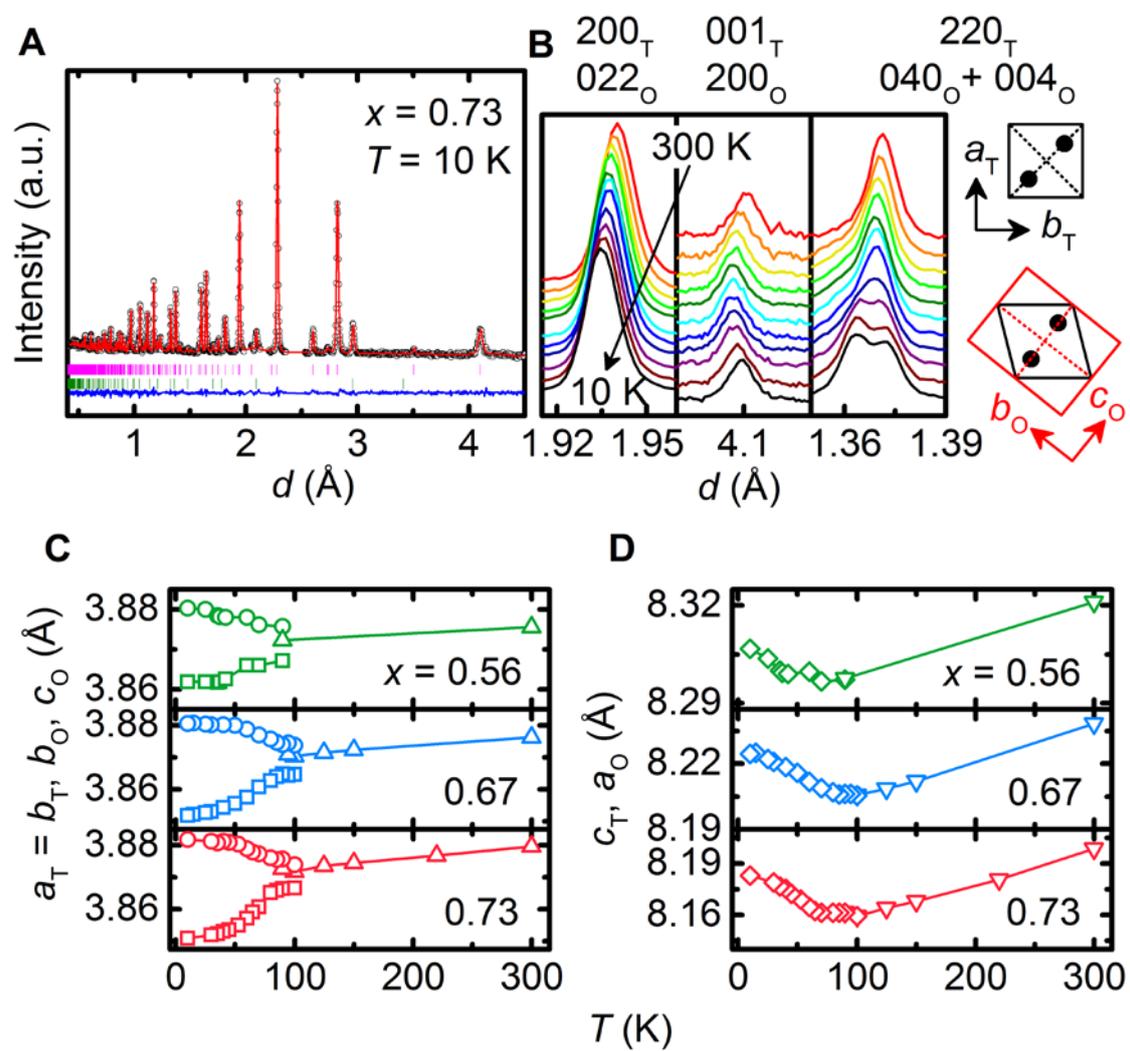

Figure 1

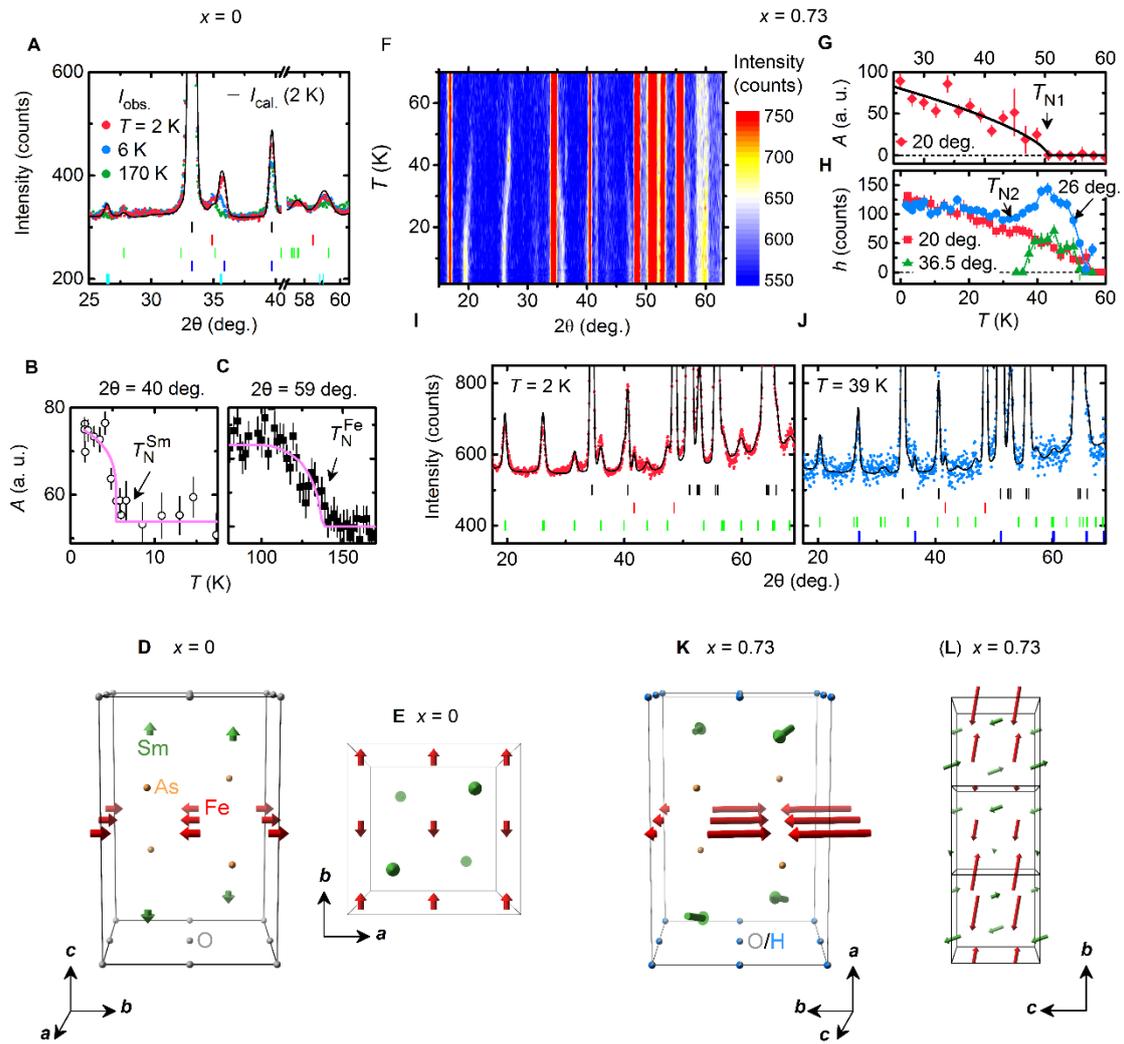

Figure 2

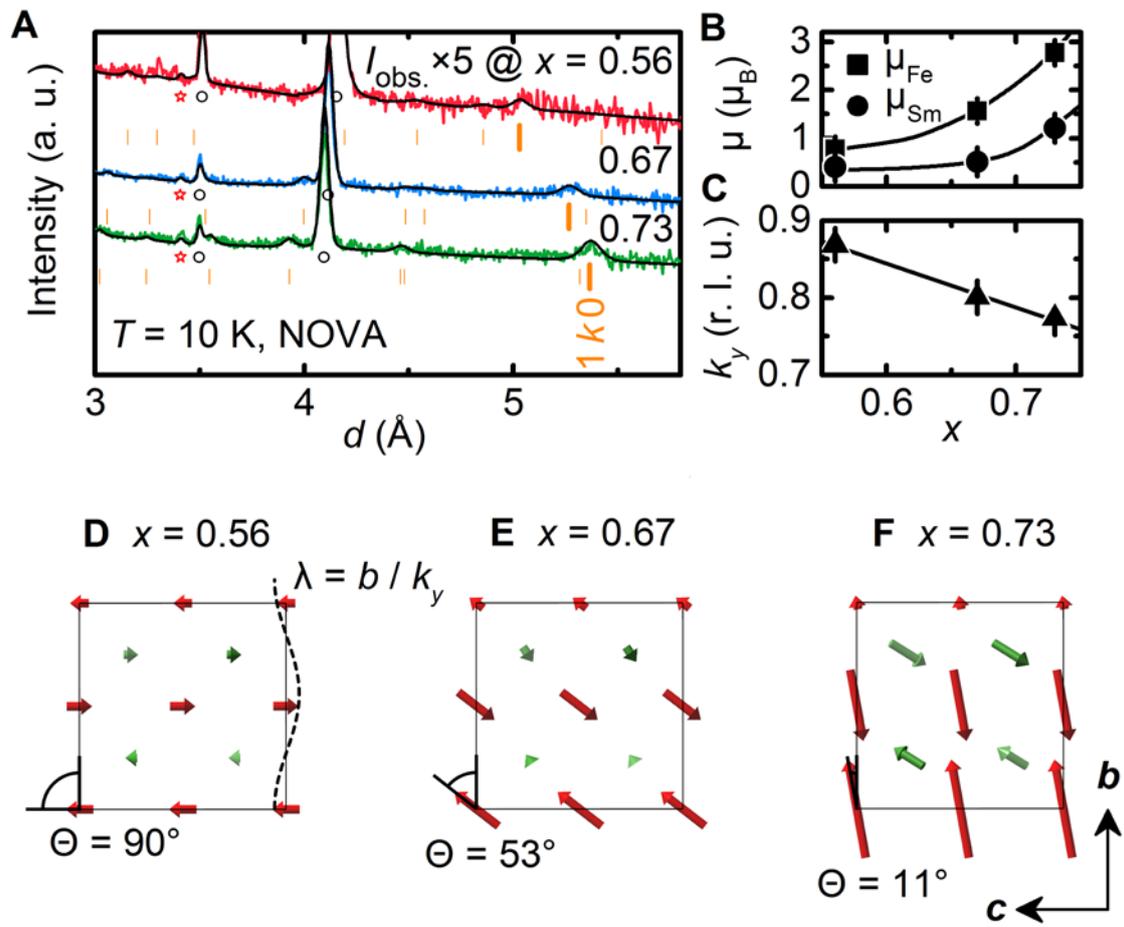

Figure 3

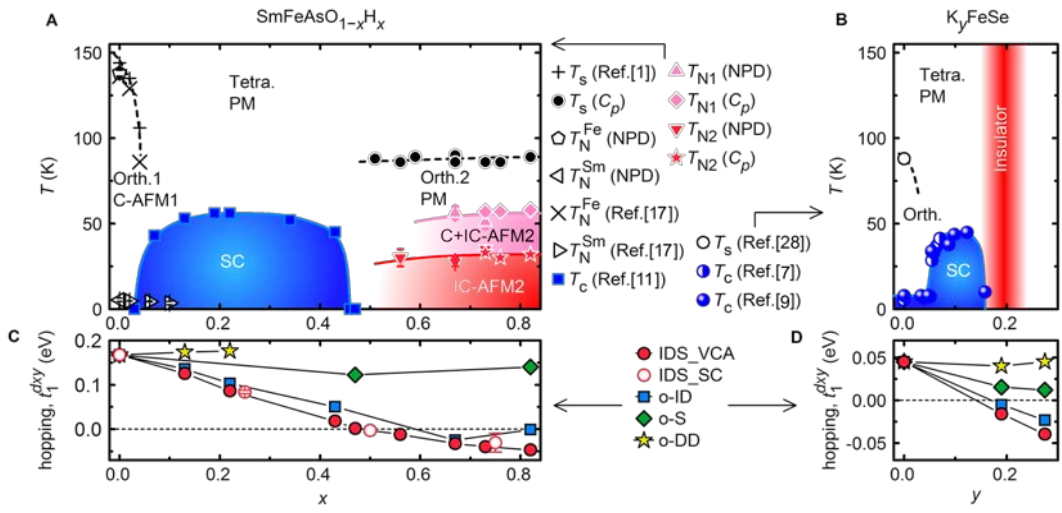

Figure 4

S. Iimura *et al.*,

**Figure Legends**

Fig. 1 Low temperature crystal structure of electron-overdoped $^{154}$SmFeAsO$_{1-x}$D$_x$
(A) NPD profile of $^{154}$SmFeAsO$_{0.27}$D$_{0.73}$ collected on NOVA at $T = 10$ K. Black circle and red solid lines denote observed and calculated intensities, respectively, while the blue solid line gives the difference between them. Pink and green tick marks below the diffraction patterns indicated predicted peak positions from the nuclear structures of $^{154}$SmFeAsO$_{0.27}$D$_{0.73}$ (89wt%) in space group *Aem*2 and a secondary phase $^{154}$SmAs (11wt%), respectively. The reliability factors for the fitting are $Rwp = 2.14\%$, $Rp = 1.95\%$, and $\chi^2 = 1.92$. (B) Temperature dependence of tetragonal 200, 001, and 220 reflections. The former two peaks are indexed as 022 and 200, respectively, with the orthorhombic *Aem*2 structure, while the later 220 reflection splits into 040 and 004 as shown above the Fig.1B. On the right size of Fig.1B we show the unit cells and corresponding crystallographic axes of the tetragonal and orthorhombic phases, where the black and red dashed lines denote the tetragonal 220 and orthorhombic 040&004 planes, respectively. (C), (D). Temperature dependence of the lattice constants of $^{154}$SmFeAsO$_{1-x}$D$_x$ ($x = 0.56$, 0.67, and 0.73). The open circle, square, and diamond denote the *b*-, *c*-, and *a*-axis lengths of the orthorhombic structure ($b_O$, $c_O$ and $a_O$, respectively), while the open triangle and inverse triangle mark the *a*- and *c*-axis lengths of tetragonal structure ($a_T$, and $c_T$, respectively). In the orthorhombic phase, the values of $b_O$ and $c_O$ are divided by √2 for easier comparison between the cell parameters of the two phases. The cell settings are related through the equations $\boldsymbol{a}_O = \boldsymbol{c}_T$, $\boldsymbol{b}_O = \boldsymbol{a}_T + \boldsymbol{b}_T$, and $\boldsymbol{c}_O = -\boldsymbol{a}_T + \boldsymbol{b}_T$.

Fig. 2 Long range AFM orderings patterns in $^{154}$SmFeAsO and $^{154}$SmFeAsO$_{0.27}$D$_{0.73}$ (A) Diffraction profiles measured for $^{154}$SmFeAsO at $T$ = 2 (red), 6 (blue), and 170 K (green) collected on D20. Black solid line denotes the calculated pattern at $T$ = 2 K. Black, red, green, blue and light-blue tick marks below the diffraction patterns represent predicted peak positions at $T$ = 2 K from respectively the nuclear structures of $^{154}$SmFeAsO (*Cmme*), FeAs, $^{154}$Sm$_2$O$_3$, the magnetic structure of Sm sublattice with $k$ = (0, 0, 0) and magnetic structure of the Fe sublattice with $k$ = (0, 1, 0.5). (B), (C) Temperature dependence of peak areas at 2θ = 40 (b) and 59 deg. (c). The fitted curves shown as solid lines are calculated using a simple power law $f(T) = A(T_N − T)^\beta + B$ (B), and a Brillouin function (C). The Neel temperatures determined from the fit are $T_N^{Sm}$ = 5.4(9) K and $T_N^{Fe}$ = 137(2) K. (D) Commensurate magnetic structure of $^{154}$SmFeAsO at $T$ = 2 K. Green, red, yellow, and grey spheres denote $^{154}$Sm, Fe, As, and O atoms, respectively. The *k*-vectors of Fe- and Sm- spin orientations are (0, 1, 0.5) and (0, 0, 0), respectively, indexed with reference to the conventional setting of the *C*-centered orthorhombic *Cmme* cell. The spins on Fe- and Sm-atoms are shown with red and green arrows, respectively. (E) Top view of the magnetic structure of $^{154}$SmFeAsO. (F) Thermodiffractogram of $^{154}$SmFeAsO$_{0.37}$D$_{0.73}$. (G) Temperature dependence of area of peak at 2θ = 20 deg. (red diamonds). The black solid line plots the best-fit function $f(T)$ = $A(T_{N1} − T)^\beta + B$ with $\beta$ = 0.33(3) and $T_{N1}$ = 51(4). (H) Temperature dependence of height of peaks at 2θ = 20 (red squares), 26 (blue circles), and 36.5 deg. (green triangles). (I) Diffraction profile of $^{154}$SmFeAsO$_{0.37}$D$_{0.73}$ at $T$ = 2 K. The black solid line denotes the calculated diffraction pattern at $T$ = 2 K, while the black, red, and green tick marks represent peak positions for the nuclear structures of $^{154}$SmFeAsO$_{0.37}$D$_{0.73}$ (*Aem*2), $^{154}$SmAs, and the magnetic structure with $k$ = (0, 0.773, 0), respectively. (J) Diffraction profile of $^{154}$SmFeAsO$_{0.37}$D$_{0.73}$ at $T$ = 39 K. The black solid line denotes the calculated pattern at $T$ = 39 K, while the black, red, green, and blue tick marks indicate the predicted peak positions for nuclear structures of $^{154}$SmFeAsO$_{0.37}$D$_{0.73}$ (*Aem*2), $^{154}$SmAs, the incommensurate magnetic structure with $k$ = (0, 0.800, 0), and the commensurate one with $k$ = (0.5, 1, 0) respectively. (K) Incommensurate magnetic structure of $^{154}$SmFeAsO$_{0.27}$D$_{0.73}$ at $T$ = 2 K. The blue color at the O positions represents the 73%

occupation of the site by D. The Fe- and Sm-spins have the same *k* vector, (0, 0.773, 0), indexed with the conventional setting of the *A*-centered orthorhombic *Aem*2 cell. The magnitude of the magnetic moments on Fe- and Sm-sites are 2.73(6)$\mu_B$/Fe and 1.7(2) $\mu_B$/Sm, respectively. (L) Top view of the magnetic structure of $^{154}$SmFeAsO$_{0.27}$D$_{0.73}$.

S. Iimura *et al*.,

Fig. 3 *x*-dependence of the heavily electron-doped AFM structure at $T = 10$ K.

(A) Diffraction patterns at $T = 10$ K collected by NOVA on samples with $x = 0.56$ (red), 0.67 (blue), and 0.73 (green). The diffraction pattern for $x = 0.56$ was multiplied by 5 for clarity. Black open circles, red stars, and orange bars below the patterns represent the peak positions from the nuclear structures of $^{154}$SmFeAsO$_{1-x}$D$_x$, $^{154}$SmAs, and the incommensurate magnetic structures with $\boldsymbol{k} = (0, k_y, 0)$, respectively. The bold orange bars denote reflection positions of the type $hkl = 1k_y0$. Black solid lines give the calculated patterns derived from Rietveld analysis. (B) *x*-dependence of the magnetic moments on the Fe (square) and Sm (circle) sites at $T = 10$ K. (C) *x*-dependence of $k_y$. Solid lines are draw as guides for the eye. (D)-(F) Top view of the AFM structures at $x = 0.56$ (D), 0.67 (E), and 0.73(F). Red and green arrows denote spins on the Fe and Sm-sites, respectively. At $x = 0.56$, the orientations of the Fe-spins create a transverse spin density wave, where the direction of the spin is perpendicular to $\boldsymbol{k}$ ($\Theta = 90°$). A dashed line shows the modulation of the magnitude of Fe spins with a wave-length $\lambda = b/k_y$. As *x* increases, the spins gradually pivot around *a*-axis, and the magnetic structure of the Fe sublattice at $x = 0.73$ approximately forms a longitudinal spin density wave where the spins are parallel to $\boldsymbol{k}$ ($\Theta = 0°$).

Fig. 4 Electronic phase diagrams of SmFeAsO$_{1-x}$H$_x$ and K$_y$FeSe.

(A) The superconducting, structural, and magnetic phase diagram determined from our NPD and $C_p$ measurements on SmFeAsO$_{1-x}$H$_x$ ($x$ = 0, 0.51, 0.59, 0.67, 0.76, 0.82) and $^{154}$SmFeAsO$_{1-x}$D$_x$ ($x$ = 0.56, 0.67, 0.73) samples. Black plus signs indicate the onset temperature of the *P*4/*nmm* to *Cmme* (Orth.1) transition adopted from Ref.[17]. Black filled circles denote the structural transition temperatures to form the *Aem*2 (Orth.2) phase, as determined by $C_p$ measurements. Black open pentagons represent the $T_N$ for the Fe sublattice's spin ordering (AFM1) determined from neutron diffraction, and the points designated by multiplication signs are from Ref. [17]. Blue filled squares show $T_c$ values taken from Ref.[11]. Pink triangles and diamonds represent transition temperatures ($T_{N1}$) for the transformation of the paramagnetic tetragonal phase to the multi-***k*** AFM (C+IC-AFM2 with the both commensurate *k*-vector ***k*** = (0.5, 1, 0) and incommensurate *k*-vector ***k*** = (0, $k_y$, 0)) phase determined from NPD and $C_p$ measurements, respectively. Red inverse triangles and stars give transition temperatures ($T_{N2}$) for the formation of the incommensurate structure with ***k*** = (0, $k_y$, 0) (IC-AFM2) determined from NPD and $C_p$ measurements, respectively. The solid and dashed lines are provided as guides for the eye. (B) The phase diagram of K$_y$FeSe. Black open circles indicate a structural transition from the *P*4/*nmm* phase to the *Cmme* one [28]. Blue open and filled squares show $T_c$ values taken from Ref.s [7] and [9], respectively. The red area at $x$ = 0.2 denote an insulating phase confirmed by ARPES measurements [7]. (C) Doping dependence of the $t_1^{dxy}$ in SmFeAsO$_{1-x}$H$_x$. The red filled circles give the $t_1^{dxy}$ values calculated using the VCA (see main text) performed on the experimental, ambient temperature crystal structure. The red open circles were obtained using a supercell in which the oxygen-site in a 1×1×1 (at $x$ = 0 and 0.5) or √2×√2×1 (at $x$ = 0.25 and 0.75) cell is partially substituted by hydrogen. Blue filled squares and yellow stars indicate value calculated using the VCA for the oxygen ($Z$ = 8 + $x$) and iron ($Z$ = 26 + $x$) sites, respectively, but with atomic positions fixed to those of nondoped SmFeAsO. Green diamonds denote the $t_1^{dxy}$ values calculated by fixing the chemical composition to SmFeAsO but changing the crystal structure according to the experimental trends. (D) Doping dependence of the $t_1^{dxy}$ values in K$_y$FeSe. The effect of indirect electron doping was approximated using the VCA for the K site ($Z$ = 18 + $y$). Blue filled squares and yellow stars give points calculated using the VCA, but with the local structure around iron sites fixed to those of nondoped FeSe. The crystal structures of the non-doped and doped phases were adopted from Ref.s [29] and [30], respectively. We note that the crystal structures of doped K$_y$FeSe with $y$ = 0.19 and 0.275 have no vacancies in the Fe-site, so the calculated $t_1^{dxy}$ values for K$_y$FeSe should reflect the electronic properties of K-coated bulk FeSe shown in Fig.4B.

# SI Appendix

# A large moment antiferromagnetic order in overdoped high-$T_\mathrm{c}$ superconductor $^{154}$SmFeAsO$_{1-x}$D$_x$

Soshi Iimura[1], Hiroshi Okanishi[1], Satoru Matsuishi[2], Haruhiro Hiraka[3], Takashi Honda[4], Kazutaka Ikeda[4], Thomas C. Hansen[5], Toshiya Otomo[4,6], and Hideo Hosono[1,2,*]

[1]Laboratory for Materials and Structures, Tokyo Institute of Technology, Yokohama 226-8503, Japan

[2]Research Center for Element Strategy, Tokyo Institute of Technology, Yokohama 226-8503, Japan

[3]Graduate School of Science and Engineering, Ibaraki University, Mito, Ibaraki 310-8512, Japan

[4]Institute of Materials Structure Science, High Energy Accelerator Research Organization (KEK), Tsukuba, Ibaraki 305-0801, Japan

[5]Institut Laue-Langevin, 6 rue Jules Horowitz, BP 156, 38042 Grenoble Cedex 9, France

[6]Department of Materials Structure Science, The Graduate University for Advanced Studies, Tsukuba, Ibaraki 305-0801, Japan

**Sample preparations**

Polycrystalline $^{154}$SmFeAsO$_{1-x}$D$_x$ with $x = 0.56, 0.67, 0.73$ and SmFeAsO$_{1-x}$H$_x$ with $x = 0.51, 0.59, 0.67, 0.76, 0.82$ were synthesized by a solid-state reaction at 1573 K under pressure of 5 GPa using a belt-type high pressure apparatus. The non-doped sample $^{154}$SmFeAsO was prepared at 1373 K under ambient pressure. The starting materials except for the $^{154}$Sm$_2$O$_3$, $^{154}$SmAs, and $^{154}$SmD$_2$ were prepared as reported previously[1]. The commercially available $^{154}$Sm$_2$O$_3$ (ATOX Co., Ltd. 98.9%) was used after dehydration by heating at 1473 K for 10 h. To obtain the $^{154}$SmD$_2$ from the $^{154}$Sm$_2$O$_3$, we first reduced the $^{154}$Sm$_2$O$_3$, as follows[2]. Figure S1A shows a schematic of the apparatus for preparation of the $^{154}$Sm metal. Lanthanum turnings 100% excess were intimately mixed with the $^{154}$Sm$_2$O$_3$, and the mixture loaded inside a tantalum crucible was heated at 1200°C for 24 h in a vacuum of 10$^{-2}$ Pa to effect the reaction: $^{154}$Sm$_2$O$_3$ + 4La → 2$^{154}$Sm + La$_2$O$_3$ + 2La. Because of the lower vapor pressure of Sm than that of La, the volatile $^{154}$Sm metal were recovered as black condensates on the wall of the copper condenser located in the upper part of the crucible. The condensates composed of $^{154}$Sm metal, $^{154}$SmO, and $^{154}$Sm$_2$O (See Fig.S1B) were heated at 673 K under deuterium gas with 0.9 MPa, and finally we obtained $^{154}$SmD$_{2-3}$. The X-ray diffraction (XRD) pattern of $^{154}$SmD$_{2-3}$ is shown in Fig.1C. We initially charged 5.8020 g of the mixture of $^{154}$Sm$_2$O$_3$ + 4La, and finally obtained 1.8895 g of condensate. If we assume that the condensate is composed of the pure $^{154}$Sm metal, the yield exceeds 96%. The $^{154}$SmAs was prepared by a reaction equation: $^{154}$Sm$_2$O$_3$ + (1+$\delta$)Ca$_3$As$_2$ → 2$^{154}$SmAs + (3−$\delta$)CaO + $\delta$Ca$_4$As$_2$O under heating at 1023 K for 20 h. The excess Ca$_2$As$_3$ ($\delta \sim 0.2$) was necessary to shift the equilibrium to the right side of the equation. The resultant mixture was added in the NH$_4$Cl-methanol solution with 0.1 mol/L, and the suspension was stirred for 1 h to dissolve the CaO and Ca$_4$As$_2$O (See Fig.S1D). The $^{154}$SmAs was collected by centrifugation at 1000 rpm for 0.5 h. The chemical compositions of the $^{154}$SmFeAsO$_{1-x}$D$_x$ and SmFeAsO$_{1-x}$H$_x$, except for hydrogen and deuterium, were measured by an electron probe micro-analyzer (JEOL JXA-8530F). The hydrogen and deuterium contents were determined by the thermal gas desorption spectrometry (ESCO TDS-1400TV) using H$_2$- and D$_2$-standard gases.

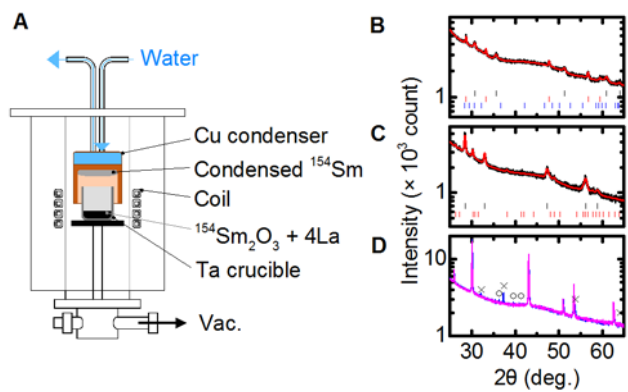

Fig. S1 Preparation of $^{154}$SmD$_2$ and $^{154}$SmAs. (A) Apparatus for preparation of $^{154}$Sm metal. (B) XRD pattern of a black condensate. Black, red, and blue tick marks represent the positions of reflection from $^{154}$SmO, $^{154}$Sm$_2$O, and $^{154}$Sm metal. (C) XRD pattern of $^{154}$SmD$_{2\text{-}3}$. Black and red tick marks denote the cubic $^{154}$SmD$_2$ and trigonal $^{154}$SmD$_3$. (D) XRD patterns of $^{154}$SmAs before and after methanol treatment. Blue and pink lines are the diffraction patterns before and after methanol treatment. The multiplication and circle signs designate the reflections from CaO and Ca$_4$As$_2$O. They disappear after methanol treatment.

**Determination of magnetic structures**

The NPD measurements were performed on the neutron total scattering spectrometer, NOVA, installed in Japan Proton Accelerator Research Complex (J-PARC) and the high flex powder diffractometer D20 at Institut Laue-Langevin (Grenobale France). For the experiment on NOVA, the crystal structure refinement was performed by the 90° detector bank with a $Q$ range of $10 < Q \leq 820$ nm$^{-1}$ and a higher resolution in a momentum space (0.6%). For the detection of the small magnetic Bragg peaks, the 45° detector bank with a $Q$ range of $4 < Q \leq 500$ nm$^{-1}$ and the lower resolution (1.2%) was also used. For the experiment on D20, a 2.41 Å wavelength from the (002) reflection of a highly oriented pyrolithic graphite was used. The samples were mounted in an enclosed cylindrical vanadium cell with a diameter of 3mm. The NPD data collected from the NOVA and D20 were analyzed using the FullPROF code[3]. To determine magnetic structures, we first searched the propagation vector ($k$-vector) using the program SARAh Refine within the Brillouin zone of crystal structure just before its magnetic transition[4,5]. Secondly, the representational analysis to determine the symmetry allowed magnetic structures was carried out by the program SARAh coRepresentational Analysis and BasIreps.

**The case for *x* = 0**

The representational analysis that tells us symmetrically adopted magnetic structures need (*i*) space group of nuclear just before magnetic transition, (*ii*) *k*-vector, and (*iii*) position of magnetic atoms[6]. As for the $^{154}$SmFeAsO, the low temperature crystal structure has already been reported; the space group is orthorhombic *Cmme*, and the sites of Sm and Fe are 4*g* and 4*a*, respectively[7]. The *k*-vector was determined from the high intensity powder patterns collected on D20. Figure S2A shows the result of thermodiffractometry for $^{154}$SmFeAsO. Below $T_N^{Fe}$ = 137 K, three magnetic Bragg peaks appear at 2θ ~ 26.3, 35.8, and 59.1 deg.. They were indexed by commensurate $k$ = (0, 1, 0.5), which is labeled $k_{17}$ = (0.5, 0.5, 0.5) in Kovalev's notation for *Oc* lattice[8]. The additional two peaks at 2θ ~35.8, and 40 deg. develop below $T_N^{Sm}$ = 5.4 K, which are indexed by $k$ = (0, 0, 0) (See Fig. S2B). Table S1 summarizes the basis vectors for Fe- and Sm-sites calculated by representational analysis using BasIreps. The best fit was obtained using $\Gamma_3$ (0.66(5)$\psi_3$) and $\Gamma_1$ (0.62(3)$\psi_1$) for the ordering of Fe's and Sm's spins, respectively.

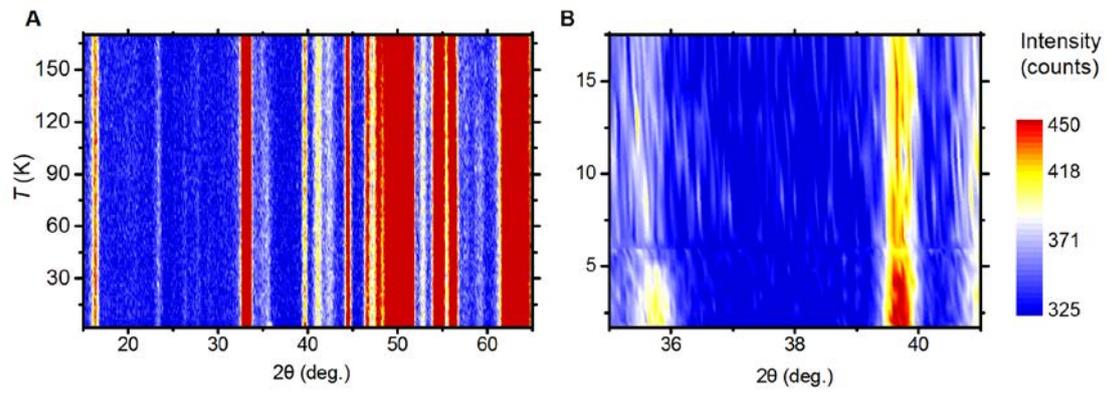

Fig. S2 Neutron diffraction on $^{154}$SmFeAsO. The thermodiffractogram collected on D20. (B) An expanded view of Fig. S2A.

Table S1 Basis vectors for the $^{154}$SmFeAsO. (A) Basis vectors for the *Cmme* structure with *k* = (0, 1, 0.5). The *k* vector is labelled $k_{17}$ = (0.5, 0.5, 0.5) in Kovalev's notation for *Oc* lattice. The decomposition of the magnetic representation for the Fe1 site (0.25, 0, 0) is $\Gamma_{Mag.}$= 1$\Gamma_1$ + 1$\Gamma_2$ + 1$\Gamma_3$ + 1$\Gamma_4$ + 1$\Gamma_5$ + 1$\Gamma_6$ + 0$\Gamma_7$ + 0$\Gamma_8$, in which all the representations are 1 dimendional. The atoms within a primitive unit call are Fe1: (0.25, 0, 0) and Fe2: (−0.25, 0, 0). (B) Basis vectors for the *Cmme* with *k* = (0, 0, 0). The *k* vector is labelled $k_{14}$ = (0, 0, 0) in Kovalev's notation for *Oc* lattice. The decomposition of the magnetic representation for the Sm1 site (0, 0.25, 0.636) is $\Gamma_{Mag.}$= 1$\Gamma_1$ + 0$\Gamma_2$ + 1$\Gamma_3$ + 1$\Gamma_4$ + 0$\Gamma_5$ + 1$\Gamma_6$ + 1$\Gamma_7$ + 1$\Gamma_8$, in which all the representations are 1 dimensional. The atoms within a primitive unit call are Sm1: (0, 0.25, 0.636) and Sm2: (0, −0.25, −0.636). The I. R. and B. V. represent Irreducible Representation and Basis Vector, respectively.

**A**

| I. R. | B. V. | Atom | B. V. components | | |
|---|---|---|---|---|---|
| | | | $m_{\|a}$ | $m_{\|b}$ | $m_{//c}$ |
| $\Gamma_1$ | $\psi_1$ | Fe1 | 1 | 0 | 0 |
| | | Fe2 | 1 | 0 | 0 |
| $\Gamma_2$ | $\psi_2$ | Fe1 | 1 | 0 | 0 |
| | | Fe2 | −1 | 0 | 0 |
| $\Gamma_3$ | $\psi_3$ | Fe1 | 0 | 1 | 0 |
| | | Fe2 | 0 | 1 | 0 |
| $\Gamma_4$ | $\psi_4$ | Fe1 | 0 | 1 | 0 |
| | | Fe2 | 0 | −1 | 0 |
| $\Gamma_5$ | $\psi_5$ | Fe1 | 0 | 0 | 1 |
| | | Fe2 | 0 | 0 | 1 |
| $\Gamma_6$ | $\psi_6$ | Fe1 | 0 | 0 | 1 |
| | | Fe2 | 0 | 0 | −1 |

**B**

| I. R. | B. V. | Atom | B. V. components | | |
|---|---|---|---|---|---|
| | | | $m_{\|a}$ | $m_{\|b}$ | $m_{//c}$ |
| $\Gamma_1$ | $\psi_1$ | Sm1 | 0 | 0 | 1 |
| | | Sm2 | 0 | 0 | −1 |
| $\Gamma_3$ | $\psi_2$ | Sm1 | 0 | 1 | 0 |
| | | Sm2 | 0 | −1 | 0 |
| $\Gamma_4$ | $\psi_3$ | Sm1 | 1 | 0 | 0 |
| | | Sm2 | −1 | 0 | 0 |
| $\Gamma_6$ | $\psi_4$ | Sm1 | 0 | 0 | 1 |
| | | Sm2 | 0 | 0 | 1 |
| $\Gamma_7$ | $\psi_5$ | Sm1 | 1 | 0 | 0 |
| | | Sm2 | 1 | 0 | 0 |
| $\Gamma_8$ | $\psi_6$ | Sm1 | 0 | 1 | 0 |
| | | Sm2 | 0 | 1 | 0 |

**The case for *x* = 0.56, 0.67, and 0.73**

To determine the space group and the position of Sm and Fe-sites of $^{154}$SmFeAsO$_{0.27}$D$_{0.73}$, we performed high resolution NPD measurements using NOVA at several temperature points below and above the structural and magnetic transitions. FIG.S3A shows the diffraction pattern at $T = 10$ K. The nuclear peaks were fitted well using a space group of *Aem*2 which is identical with the space group of LaFeAsO$_{0.49}$H$_{0.51}$[9]. The obtained structural parameters of $^{154}$SmFeAsO$_{1-x}$D$_x$ with $x = 0.56, 0.67,$ and 0.73 are summarized in Table S2. Figure S3B shows the thermodiffractogram for $^{154}$SmFeAsO$_{0.27}$D$_{0.73}$. At the lowest temperature of $T = 2$ K, we found more than 10 magnetic Bragg peaks in the range of $2\theta$ from 15 to 70 deg.. The *k*- vector was determined using SARAh Refine code to be incommensurate (0, 0.773, 0), which is labeled $k_{10} = (\mu, \mu, 0)$ in Kovalev's notation for *Oa* lattice. Table S3 summarizes the basis vectors for the $^{154}$SmFeAsO$_{0.27}$D$_{0.73}$ with the incommensurate *k*-vector. The best fit was obtained using $\Gamma_1$, and the refined coefficients of the linear combination between the $\psi_1, \psi_2,$ and $\psi_3$ are listed for each *x* of 0.56, 0.67, and 0.73 in Table S4. In the temperature range of 30 ($T_{N2}$) $< T < 51$ K ($T_{N1}$) the magnetic Bragg peaks shift according to the change of the *k*-vector from (0, 0.773, 0) to (0, 0.800, 0) on heating, and additional peaks with a commensurate *k*-vector of (0.5, 1, 0) develop at $2\theta = 26$ and 36.5 deg., which means that the single-*k* incommensurate magnetic structure with $k = (0, 0.773-0.800, 0)$ transforms into a multi-*k* structure in the temperature range. Table S5 summarizes the basis vectors for the $^{154}$SmFeAsO$_{0.27}$D$_{0.73}$ with the *k*-vector = (0.5, 1, 0). The fitting was conducted using two magnetic phases with the incommensurate *k* and the commensurate *k* of (0.5, 1, 0) at $T = 39$ K, and the best fit was obtained using $\Gamma_4$ (0.42$\psi_6$). Figure S4 illustrates the temperature variation in the magnetic structure at $x = 0.73$. Below $T_{N1} = 51(4)$, the paramagnetic spins order in the multi-*k*-magnetic structure composed of commensurate (C-AFM) and incommensurate antiferromagnetic structures (IC-AFM). The C-AFM has a same in-plane (*bc*-plane) spin configuration as observed in the AFM of LaFeAsO$_{0.49}$H$_{0.51}$. Below $T_{N2} = 30$ K, the C-AFM vanishes, and the remained IC-AFM changes *k*-vector from (0, 0.800, 0) at $T = 39$ K to (0, 0.773, 0) at $T = 2$ K as explained above.

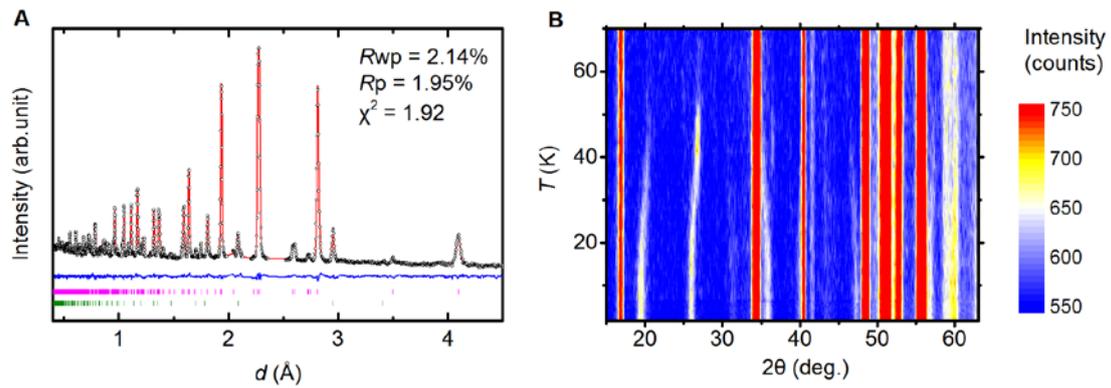

Fig. S3 Neutron diffraction on $^{154}$SmFeAsO$_{0.27}$D$_{0.73}$. (A) NPD profile of $^{154}$SmFeAsO$_{0.27}$D$_{0.73}$ collected on NOVA at $T$ = 10 K. Black circle and red solid lines denote observed and calculated intensities, respectively, and the blue solid line is a difference between them. Pink and green tick marks below the diffraction patterns represent diffraction positions from the nuclear structures of $^{154}$SmFeAsO$_{0.27}$D$_{0.73}$ with a space group of *Aem*2 and a secondary phase $^{154}$SmAs, respectively. The reliable factor for the fitting are $Rwp$ = 2.14%, $Rp$ = 1.95%, and $\chi^2$ = 1.92. (B) The thermodiffractogram collected on D20.

Table S2 Structural parameters of $^{154}$SmFeAsO$_{1-x}$D$_x$ with $x$ = 0.56, 0.67, and 0.73 at $T$ = 10 K.

| $x = 0.56$ | | | Space group : *Aem*2 | | |
|---|---|---|---|---|---|
| $a$ = 8.30659(9), $b$ = 5.48757(8), $c$ = 5.46153(7) | | | | | |
| Atom | $x$ | $y$ | $z$ | Occ. | $100 \times U_{iso}$ |
| $^{154}$Sm | 0.15566(8) | 0.25 | 0.257(3) | 1 | 0.63(1) |
| Fe | 0 | 0 | 0.0103(3) | 1 | 0.198(9) |
| As | 0.6729(8) | 0.25 | 0.252(15) | 1 | 0.34(2) |
| O | 0 | 0 | 0 | 0.44 | 1.16(6) |
| D | 0 | 0 | 0 | 0.56 | 2.07(5) |
| $x = 0.67$ | | | Space group : *Aem*2 | | |
| $a$ = 8.2245(1), $b$ = 5.48771(8), $c$ = 5.44695(8) | | | | | |
| Atom | $x$ | $y$ | $z$ | Occ. | $100 \times U_{iso}$ |
| $^{154}$Sm | 0.1594(1) | 0.25 | 0.258(3) | 1 | 0.42(3) |
| Fe | 0 | 0 | 0.011(1) | 1 | 0.27(2) |
| As | 0.6752(8) | 0.25 | 0.254(3) | 1 | 0.26(3) |
| O | 0 | 0 | 0 | 0.33 | 0.58(11) |
| D | 0 | 0 | 0 | 0.67 | 0.98(8) |
| $x = 0.73$ | | | Space group : *Aem*2 | | |
| $a$ = 8.1838(1), $b$ = 5.49038(8), $c$ = 5.44661(8) | | | | | |
| Atom | $x$ | $y$ | $z$ | Occ. | $100 \times U_{iso}$ |
| $^{154}$Sm | 0.1612(4) | 0.25 | 0.257(2) | 1 | 0.32(2) |
| Fe | 0 | 0 | 0.008(2) | 1 | 0.098(10) |
| As | 0.6759(4) | 0.25 | 0.253(2) | 1 | 0.19(2) |
| O | 0 | 0 | 0 | 0.27 | 0.79(4) |
| D | 0 | 0 | 0 | 0.73 | 1.31(6) |

Table S3 Basis vectors for the $^{154}$SmFeAsO$_{0.27}$D$_{0.73}$ with an incommensurate *k*-vector. The magnetic structure of $^{154}$SmFeAsO$_{1-x}$D$_x$ below $T = T_{N2}$ is a single-*k* structure with an incommensurate $k = (0, k_y, 0)$. The *k* vector is labelled $k_{10} = (\mu, \mu, 0)$ in Kovalev's notation for *Aem*2 (*Oa* lattice). The decomposition of the magnetic representation for the Fe1 site (0.5, 0, *z*) and Sm1 site (*x*, 0.25, *z*) are $\Gamma_{\text{Mag.}} = 3\Gamma_1 + 3\Gamma_2$, in which all the representations are 1 dimensional. The atoms within a primitive unit call are Fe1: (0.5, 0, *z*), Fe2: (0.5, 0.5, *z*), Sm1: (*x*, 0.25, *z*), and Sm2: (1−*x*, 0.75, *z*). The *ε* represents a phase factor expressed as $exp\,(-\pi i k_y)$.

**A**

| I. R. | B. V. | Atom | B. V. components | | |
|---|---|---|---|---|---|
| | | | $m_{\|a}$ | $m_{\|b}$ | $m_{//c}$ |
| $\Gamma_1$ | $\psi_1$ | Fe1 | 1 | 0 | 0 |
| | | Fe2 | −ε | 0 | 0 |
| | $\psi_2$ | Fe1 | 0 | 1 | 0 |
| | | Fe2 | −1 | 0 | 0 |
| | $\psi_3$ | Fe1 | 0 | ε | 0 |
| | | Fe2 | 0 | 0 | ε |
| $\Gamma_2$ | $\psi_4$ | Fe1 | 1 | 0 | 0 |
| | | Fe2 | ε | 0 | 0 |
| | $\psi_5$ | Fe1 | 0 | 1 | 0 |
| | | Fe2 | 0 | −ε | 0 |
| | $\psi_6$ | Fe1 | 0 | 0 | 1 |
| | | Fe2 | 0 | 0 | −ε |

**B**

| I. R. | B. V. | Atom | B. V. components | | |
|---|---|---|---|---|---|
| | | | $m_{\|a}$ | $m_{\|b}$ | $m_{//c}$ |
| $\Gamma_1$ | $\psi_1$ | Sm1 | 1 | 0 | 0 |
| | | Sm2 | −ε | 0 | 0 |
| | $\psi_2$ | Sm1 | 0 | 1 | 0 |
| | | Sm2 | −1 | 0 | 0 |
| | $\psi_3$ | Sm1 | 0 | ε | 0 |
| | | Sm2 | 0 | 0 | ε |
| $\Gamma_2$ | $\psi_4$ | Sm1 | 1 | 0 | 0 |
| | | Sm2 | ε | 0 | 0 |
| | $\psi_5$ | Sm1 | 0 | 1 | 0 |
| | | Sm2 | 0 | −ε | 0 |
| | $\psi_6$ | Sm1 | 0 | 0 | 1 |
| | | Sm2 | 0 | 0 | −ε |

Table S4 Results of Rietveld refinements for the IC-AFM phase at $T = 10$ K. The $C_1$, $C_2$, and $C_3$ are the coefficients of the linear combination between the $\psi_1$, $\psi_2$, and $\psi_3$ of $\Gamma_1$ listed in Table S3. $\Phi$ is a refined relative phase between the spins on Sm- and Fe-sites. $k_y$ is a *y*-component of the incommensurate *k*-vector.

| | $C_1 / \mu_B$ | $C_2 / \mu_B$ | $C_3 / \mu_B$ | $\Phi$ | $k_y$ |
|---|---|---|---|---|---|
| | | | $x = 0.56$ | | |
| Sm | 0 | 0 | 0.5(3) | 0 | 0.869(6) |
| Fe | 0 | 0 | 0.7(1) | 0 | |
| | | | $x = 0.67$ | | |
| Sm | 0 | 0.5(3) | 0.3(1) | 0.08(3) | 0.801(4) |
| Fe | 0 | 0.9(3) | 1.2(1) | 0 | |
| | | | $x = 0.73$ | | |
| Sm | 0.19(8) | 0.4(2) | 1.6(2) | 0.11(3) | 0.773(4) |
| Fe | 0.1(1) | 2.71(6) | 0.3(2) | 0 | |

Table S5 Basis vectors for the $^{154}$SmFeAsO$_{0.27}$D$_{0.73}$ with a commensurate *k*-vector. Basis vectors for the *Aem*2 structure with $k$ = (0.5, 1, 0). The *k* vector is labelled $k_{17}$ = (0.5, 0.5, 0.5) in Kovalev's notation for *Oa* lattice. The decomposition of the magnetic representation for the Fe1 site (0.5, 0, *z*) is $\Gamma_{Mag.}$= 2$\Gamma_1$ + 2$\Gamma_2$ + 1$\Gamma_3$ + 1$\Gamma_4$, in which all the representations are 1 dimendional. The atoms within a primitive unit call are Fe1: (0.5, 0, *z*) and Fe2: (0.5, 0.5, *z*).

| I. R. | B. V. | Atom | B. V. components | | |
|---|---|---|---|---|---|
| | | | $m_{\|a}$ | $m_{\|b}$ | $m_{\|/c}$ |
| $\Gamma_1$ | $\psi_1$ | Fe1 | 1 | 0 | 0 |
| | | Fe2 | 1 | 0 | 0 |
| | $\psi_2$ | Fe1 | 0 | 1 | 0 |
| | | Fe2 | 0 | −1 | 0 |
| $\Gamma_2$ | $\psi_3$ | Fe1 | 1 | 0 | 0 |
| | | Fe2 | −1 | 0 | 0 |
| | $\psi_4$ | Fe1 | 0 | 1 | 0 |
| | | Fe2 | 0 | 1 | 0 |
| $\Gamma_3$ | $\psi_5$ | Fe1 | 0 | 0 | 1 |
| | | Fe2 | 0 | 0 | 1 |
| $\Gamma_4$ | $\psi_6$ | Fe1 | 0 | 0 | 1 |
| | | Fe2 | 0 | 0 | −1 |

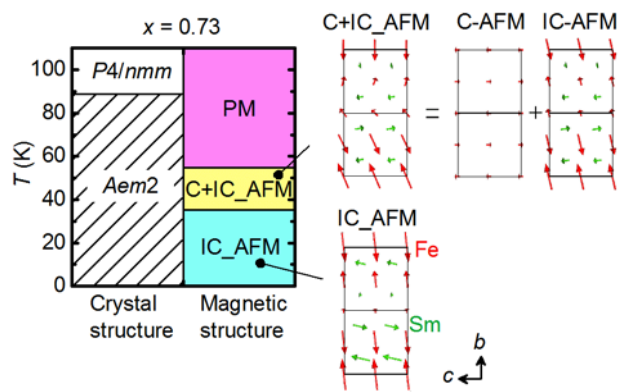

Fig. S4 Temperature variation in the crystal and magnetic structures of $^{154}$SmFeAsO$_{0.27}$D$_{0.73}$. Left figure represents the temperature variation in the crystal and magnetic structures of $^{154}$SmFeAsO$_{0.27}$D$_{0.73}$, while in the right, we illustrates the magnetic structures of C+IC_AFM and IC_AFM.

**Physical properties**

**Heat capacity**

Figure S5A and S5B show a temperature dependence of heat capacity ($C_p$) of $^{154}$SmFeAsO. The $C_p$ ($T$) curve of $^{154}$SmFeAsO has two peaks at $T \sim$ 140 and $\sim$ 5 K. The former is attributed to a magneto-structural transition from the paramagnetic tetragonal to antiferromagnetic orthorhombic as observed in most iron oxypnictides[10], while the later transition at $T \sim$ 5 K is to the AFM transition of Sm[7]. In Fig.S5C and S5D, we plot the $C_p/T$ v.s. $T$ of over-doped samples with $x \geq 0.51$. A small peak at $T \sim$ 90 K observed in all 8 samples is from the structural transition from tetragonal *P*4/*nmm* to orhothombic *Aem*2. In addition to this transition, we also found other two anomalies at $T \sim$ 30 and 55 K and at $x = $ 0.73, 0.76, and 0.82, which are obvious in the temperature derivative of the $C_p$ shown in Fig.S5F-H. The higher temperature anomaly is located at $T \sim$ 57, 57, and 58 K for the samples with $x = $ 0.73, 0.76, and 0.82, respectively, and the lower is at 33, 30, and 32 K for the samples with $x = $ 0.73, 0.76, and 0.82, respectively which are determined from the intersection of the two extrapolated lines. As for the sample with $x = $ 0.73, the higher transition temperature of 57 K coincides well with the $T_{N1}$ of 51(4) K, while the lower of 30 K does well with the $T_{N2}$ of 35 K, thus indicating that those anomalies are from these magnetic transitions. Unlike the non-doped and lightly doped samples, a clear peak originating the AFM transition of Sm cannot be observed by the $C_p$ measurement in $x > 0.51$.

S. Iimura *et al.*,

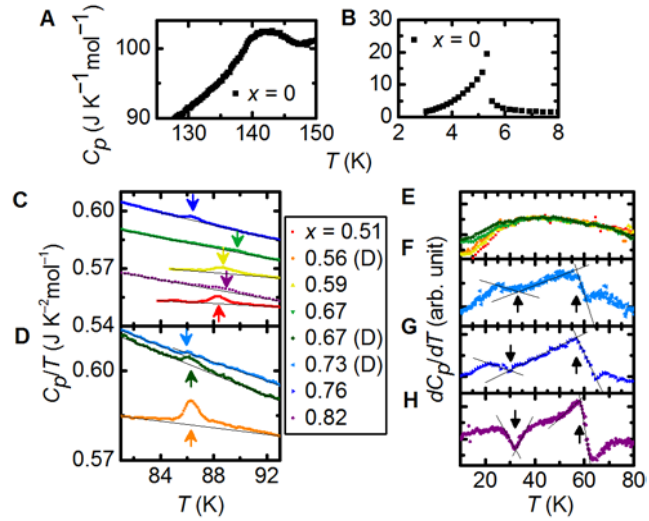

Fig. S5 Results of heat capacity measurements on SmFeAsO$_{1-x}$H$_x$ and $^{154}$SmFeAsO$_{1-x}$D$_x$. (A, B) $C_p$-$T$ curves of $^{154}$SmFeAsO. (C, D) Temperature dependence of heat capacity ($C_p/T$ vs. $T$) for SmFeAsO$_{1-x}$H$_x$ (C) and $^{154}$SmFeAsO$_{1-x}$D$_x$ (D). The arrows indicate the structural transition temperature ($T_s$). Black solid lines are guide for eyes. (E-G) Temperature dependence of $dC_p(T)/dT$ at $x$ = 0.51-0.67 (E), 0.73 (F), 0.76 (G), and 0.82 (H).

**Resistivity**

The electrical resistivity, $\rho$ ($T$), was measured using a four-probe technique. Figure S6A shows temperature dependence of $\rho$ ($T$) of SmFeAsO$_{1-x}$H$_x$ and $^{154}$SmFeAsO$_{1-x}$D$_x$. All the samples show no superconducting transition, which is evident that they are overdoped samples. In the $d\rho$ ($T$)$/dT$ curves shown in Fig.S6B, some apparent kinks were observed below 100 K. In the cases at $x$ = 0.73, 0.76, and 0.82, the higher temperature anomalies at 54, 58, and 52 K coincide well with their $C_p$ anomalies corresponding to the $T_{N1}$ at 57, 57, and 58 K, while the lower ones at 33, 32, and 27 K does well with their $C_p$ anomalies corresponding to the $T_{N2}$ at 33, 30, and 32 K. In the $x$ range of $0.59 \leq x \leq 0.67$, we can see that the anomaly corresponding to the $T_{N2}$ splits into two. Moreover, at $x$ = 0.51 and 0.56, there is another kink below $T$ = 20 K. Unfortunately, our neutron diffraction and $C_p$ measurements cannot detect any phase transitions at these temperature. A more sensitive probe to the change of structure and/or magnetism, such as synchrotron x-ray diffraction and muon spin rotation measurements, may be necessary to detect them and reveal the origin.

S. Iimura *et al.*,

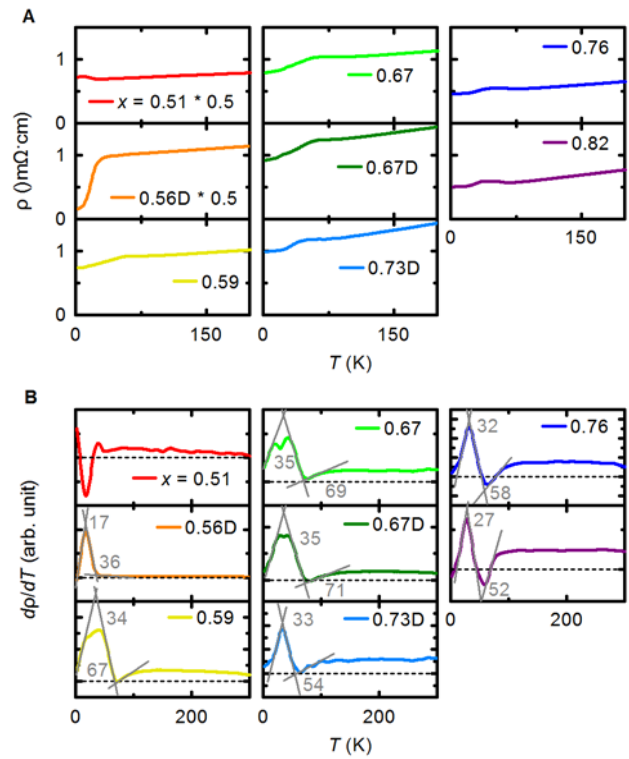

Fig. S6 Results of ρ (*T*) measurements on SmFeAsO$_{1-x}$H$_x$ and $^{154}$SmFeAsO$_{1-x}$D$_x$. (A) Temperature dependence of resistivity (ρ (*T*)). (B) The temperature differentials of ρ (*T*). To obtain the differential of *d*ρ (*T* )/*dT* , we used the Savitzky-Golay scheme. The anomalies were determined from the intersection of the two extrapolated lines.

**Theoretical calculations**

First-principles calculations were performed using the VASP code[11,12]. For the generalized gradient approximation (GGA), PBEsol formulated by the Perdew, Burke, and Ernzerhof was adopted Ref. [13]. The wave functions are expanded by plane waves up to cut-off energy of 500 eV, and $6 \times 6 \times 3$ and $4 \times 4 \times 3$ meshes were used for the $1 \times 1 \times 1$, and $\sqrt{2} \times \sqrt{2} \times 1$ tetragonal cells, respectively. To estimate the hopping parameters of Fe-$3d$ orbitals, maximally localized Wannier functions were constructed from bands derived from Fe $3d$ states within an energy window from $E - E_\mathrm{F} \sim -2$ to $\sim 2$ eV using wannier90 code[14].

As we mentioned in the text, we performed 4 kinds of calculations, that is, IDS_VCA, o-ID, o-S, and o-DD for the SmFeAsO$_{1-x}$H$_x$ and K$_y$FeSe. As for the SmFeAsO$_{1-x}$H$_x$, we also calculated the band structure using a supercell containing H (IDS_SC). The IDS_VCA model includes the effects of experimental structural changes and the indirect electron doping. The structural data of SmFeAsO$_{1-x}$H$_x$ at ambient temperature were determined by our XRD measurements, while those of K$_y$FeSe were taken from Ref.[15] for $y = 0$ and [16] for $y > 0$ (see Fig. S7). The effect of indirect electron doping is included using a virtual crystal approximation (VCA). For the SmFeAsO$_{1-x}$H$_x$, we applied the VCA on oxygen site by mixing the pseudo potentials of oxygen and fluorine, while for the K$_y$FeSe, we use the VCA on the potassium site by mixing those of K and Ar. The IDS_SC also includes both the effects of structural change and indirect doping. However, here, we use a supercell to approximate the indirect electron doping by substituting hydrogen in oxygen site. For the calculations of SmFeAsO$_{0.75}$H$_{0.25}$, SmFeAsO$_{0.5}$H$_{0.5}$, and SmFeAsO$_{0.25}$H$_{0.75}$, the $\sqrt{2} \times \sqrt{2} \times 1$, $1 \times 1 \times 1$, and $\sqrt{2} \times \sqrt{2} \times 1$ cells were used, respectively. The o-ID model includes only the effect of indirect electron doping using the VCA. For the calculations of SmFeAsO$_{1-x}$H$_x$, we used the structure of SmFeAsO, while for the K$_y$FeSe, we changed the Fe-Fe and Fe-Se distances of K$_y$FeSe with $y > 0$ to those of FeSe. The o-S model includes only the effect of structural change, where we adopted the same structures used in IDS_VCA model. Finally, the o-DD model includes only the effect of direct electron doping by using the VCA mixing the pseudo potentials of Fe and Co. Here, we adopted the same structures used in o-ID model.

S. Iimura *et al.*,

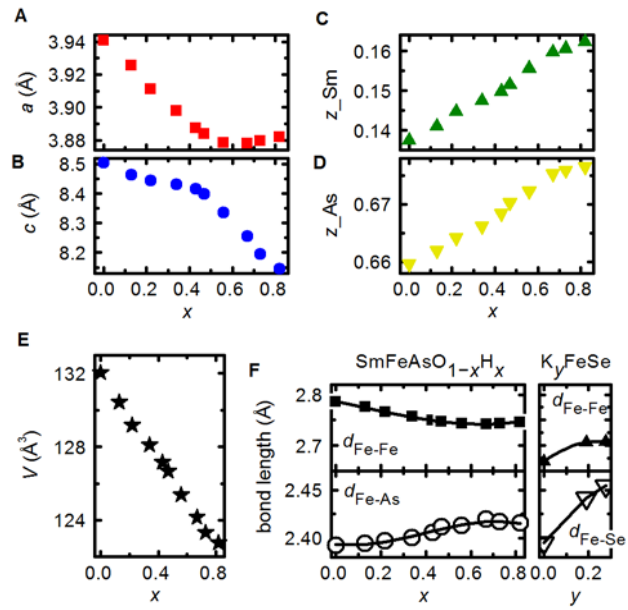

Fig. S7 Structural parameters of SmFeAsO$_{1-x}$H$_x$ and K$_y$FeSe. (A-E) *x* dependence of structural parameters of SmFeAsO$_{1-x}$H$_x$ at ambient temperature. (F) Fe-Fe and Fe-As or -Se bond lengths of SmFeAsO$_{1-x}$H$_x$ (left two panels) and K$_y$FeSe (right two panels) [15,16].